\documentclass[11pt]{article}

\usepackage[a4paper, margin=1in]{geometry}

\usepackage{graphicx}
\usepackage{hyperref}
\usepackage{amsmath,amssymb}
\usepackage{siunitx}
\usepackage{authblk}
\usepackage{multicol}
\usepackage{bibentry}

\usepackage{titlesec}
\titleformat*{\section}{\large\bfseries\sffamily}
\titleformat*{\subsection}{\bfseries\sffamily}

\def\targ#1{{\bf\boldmath#1}}

\newif\ifshortversion
\ifshortversion
\def\maybecite#1{\relax}%

\def\mayberef{\noindent\textit{A version of this white paper with a more complete bibliography is available on \href{https://arxiv.org/abs/astro-ph/0608305}{arXiv:2512.9999}}}%
\else
\let\maybecite\cite
\def\mayberef{\noindent\textit{This version can be posted on arXiv, while a version with a shorter bibliography and no abstract is submitted to ESO.}}%
\fi

\title{\bf Stellar physics at sub-nanoradian angular resolution}

\author[1]{P.~Batista}
\author[2]{J.~Biteau}
\author[3]{C.~Carlile}
\author[4]{J.~Cortina}
\author[5]{D. Della Volpe}
\author[3]{D.~Dravins}
\author[6]{M.~Fiori}
\author[1]{S.~Funk}
\author[7]{W.~Guerin}
\author[4]{T.~Hassan}
\author[1]{C.~Ingenhütt}
\author[8]{I.~Jim\'enez~Mart\'inez}
\author[7]{R.~Kaiser}
\author[5]{G.~Koziol}
\author[9]{O.~Lai}
\author[2]{Q.~Luce}
\author[5]{E.~Lyard}
\author[8]{R.~Mirzoyan}
\author[1]{A.~W.~M.~Mitchell}
\author[10]{A.~Nomerotski}
\author[5]{N.~Produit}
\author[5]{A.~Raiola}
\author[11]{P.~Saha}
\author[8]{T.~Schweizer}
\author[5]{V.~Sliusar}
\author[6]{A.~Spolon}
\author[11]{L.~Stanic}
\author[5]{R.~Walter}
\author[6]{L.~Zampieri}
\author[1]{J.~von~Zanthier}
\author[7]{A.~Zmija}
\author[12]{J.~Aufdenberg}
\author[13]{C.~F.~Bender}
\author[14]{J.~J.~Hermes}
\author[15]{D.~Kieda}
\author[16]{M.~Lisa}
\author[17]{A.~Mueninghoff}
\author[18]{K.~N.~Rai}
\author[19]{K.~van~Tilburg}

\affil[1]{Friedrich-Alexander-Universität Erlangen-Nürnberg, Germany}
\affil[2]{Université Paris-Saclay, IJCLab, France}
\affil[3]{Lund University, Sweden}
\affil[4]{CIEMAT, Spain}
\affil[5]{University of Geneva, Switzerland}
\affil[6]{INAF-Osservatorio Astronomico di Padova, Italy}
\affil[7]{Universit\'e C\^ote d’Azur, CNRS, Institut de Physique de Nice, 06200 Nice, France}
\affil[8]{Max Planck Institute for Physics, Germany}
\affil[9]{Universit\'e C\^ote d’Azur, Observatoire de la C\^ote d’Azur, CNRS, Laboratoire Lagrange, 06000 Nice, France}
\affil[10]{Czech Technical University in Prague, Czechia}
\affil[11]{University of Zurich, Switzerland}
\affil[12]{Embry-Riddle Aeronautical University, USA}
\affil[13]{University of Arizona, USA}
\affil[14]{Boston University, USA}
\affil[15]{University of Utah, USA}
\affil[16]{Ohio State University, USA}
\affil[17]{Stonybrook University, USA}
\affil[18]{Aryabhatta Research Institute of Observational Sciences, India}
\affil[19]{New York University, USA}
\date{}

\begin{document}

\maketitle

\begin{abstract}
\noindent Many stars --- if they could be imaged with enough angular resolution --- would exhibit features expected from theory but not possible to extract from spectra.  We may group these by increasing complexity as follows.  First, smooth variations in brightness across the surface, resembling solar limb darkening but much more prominent and involving more processes in stars with fast spin or external tides.  Next, there are periodic features: not only oscillations, but also convective cells and starspots, which appear to transit across a star as its spins, and exoplanets that really do transit across the star.  Then, there are transients like flares.

Current optical interferometers provide synthetic apertures of a few hundred metres and angular resolutions down to about nanoradian ($\simeq 0.2\,$milliarcsecond), enough to resolve some of the above features on the nearest upper main-sequence stars, giants and supergiants.  Ongoing projects aims to km-scale synthetic apertures, enough to measure the radius of the nearest white dwarf.  In this White Paper we briefly discuss what could be observed with synthetic apertures over $\sim20\,$km --- resolving detail on white dwarfs at the level currently possible on supergiants.
\end{abstract}

\thispagestyle{empty}

\clearpage

\setcounter{page}{1}

\section{The angular-resolution frontier for stars}

Imagine alien astronomers observing the Sun from parsecs away, using
spectroscopic telescopes but lacking imagers.  Such aliens could estimate the mass and age of the Sun, infer much about its internal structure from asteroseismic modes, and deduce the presence of several planets.  But they would miss solar granules, the solar dynamo, and coronal heating.  In other words, angular resolution reveals a great deal even on a star with a modest magnetic field and surface convection, slow spin, and insignificant oscillations, outflows, or external tides.  In stars less placid than the Sun, how much more are \textit{we}
missing??

Humans are currently doing somewhat better than our hypothetical
aliens --- we now have reconstructions of surface structure on a few supergiants and upper main-sequence stars.  Recent examples are giant convective cells on \targ{$\rho$~Cassiopeiae}\cite{2024ApJ...974..113A}, and the
rotational flattening and gravity darkening of
\targ{$\gamma$~Cassiopeiae}\cite{2025arXiv250615027A}.  What if our telescopic vision became much sharper?  What if we had the level of detail on the nearest white dwarf (\targ{Sirius~B}) that we now have on \targ{Canopus}\maybecite{2021A&A...654A..19D}, the
nearest supergiant?  In this White Paper, we discuss the stellar physics we could learn, and then argue that the goal is reachable.

\begin{figure}[h]
\includegraphics[width=\linewidth]{fig/somestars_shaded}
\caption{\em The angular-resolution frontier for stars.  Plotted are the sizes and distances of examples mentioned in the text.  Colors are approximately true color, corresponding to effective temperature. The shaded regions correspond to current and proposed (synthetic) telescope apertures. IACT refers to Cherenkov telescopes in intensity-interferometry mode.}
\end{figure}

Figure~1 shows the stellar angular-resolution frontier using a selection of stars as examples.  White dwarfs are to the left, supergiants are to the right. Stars in the lower-right diagonal are currently resolvable.  Stars in the upper-left diagonal are future targets.  The solar analog and exoplanet host \targ{HD~209458} is at about the current angular-resolution limit about 200~micro-arcseconds or more concisely a nanoradian.  The EHT has (at mm-wavelengths) a resolution of 100~picoradians.  An optical interferometer with a baseline of 20~km would reach 20~picoradians.

\section{Steady processes}

The simplest application of high angular-resolution on stars is for testing mass-radius relations.  There are indirect tests of mass-radius relations through spectrophotometry, even for the smallest stellar objects\maybecite{2017ApJ...848...11B,2018MNRAS.481.2361J}.
But direct radius measurements of any star are valuable when
available\maybecite{2022MNRAS.515....1D}.  Theory tells us that
\targ{Rigel} (for example) must be supported by radiation pressure,
\targ{$\alpha$~Centauri} A \& B by gas pressure, \targ{Luhman~16} A \&
B by non-relativistic degenerate electrons, and \targ{Sirius~B} by relativistic electron degeneracy, so a large program of radius measurements would be confronting predictions of models involving diverse physical processes, including macroscopic quantum systems in the case of degenerate remnants.  Perhaps some even stranger creature such as a Thorne-{\.Z}ytkow object\maybecite{2023MNRAS.524.1692F} will reveal itself.  
Targets in multi-star systems are especially important here, because orbits provide good mass measurements. (Some examples appear in Figure~1 as stars of different sizes at exactly the same distance.)   Mass measurements from weak microlensing (demonstrated in the case of the white dwarf \targ{Stein 2051~B}\maybecite{2017Sci...356.1046S} and indirectly also giving the mass of its red-dwarf companion \targ{Stein 2051~A}) may become more common in the future, thus increasing the number of useful mass-radius targets.

Starlight does not come from a distinct surface but from an
atmosphere that emits, absorbs, and scatters with polarization at various depths.  Additional considerations are spin and external tides, which may significantly change the local surface gravity.  As a result, stars are not uniform-brightness disks on the sky.  Some stars depart more strongly from uniform brightness: \targ{Regulus} for example is an oblate fast rotator showing gravity darkening of its equatorial region\maybecite{2011ApJ...732...68C}, and also a net polarization\maybecite{2017NatAs...1..690C} which remains to be spatially resolved.  Even the traditional photometric reference star \targ{Vega} turns out under interferometry to have a very unexpected spatial brightness profile\maybecite{2006ApJ...645..664A}, as it is a nearly pole-on fast rotator.  Next-generation angular resolution will show us stellar atmospheres in novel contexts.

\section{Periodic and quasi-periodic processes}

Here we consider phenomena that are time-dependent but predictably repeating.

First, the southern classical Cepheid \targ{$\beta$~Doradus} is shown in Figure~1 to remind us of all pulsating stars, radial and non-radial, whose oscillations could be spatially resolved.

Close multistar systems (but not close enough for accretion) also fall in the periodic category.  As we remarked above, these are especially useful for testing mass-radius predictions.  In nearby binaries, regular telescopes can resolve the orbits, with interferometry needed only for radii.  In more distant systems like \targ{Spica} interferometry is needed for the orbit as well\maybecite{1971MNRAS.151..161H}.  The two \targ{Spica} stars are actually close enough that the mutual tidal fields become significant,
making this a weakly interacting binary.  Still more exotic are dwarf stars in very close binaries, which are ``verification'' gravitational-wave sources for the LISA mission. \targ{CD-$30^\circ$~11223} is an example, where resolving it optically would predict the GW polarization, thus providing a test of general relativity versus alternative-gravity theories\maybecite{2020MNRAS.498.4577B}.

Static or slowly-varying features on a stellar surface may appear, like sunspots, to transit because of spin.  We have already mentioned convective cells on stars.  These are expected from theory\maybecite{1975ApJ...195..137S} to be few but gigantic in giant stars, but much smaller and more numerous in solar-type stars.  Starspots due to metal pollution are expected even on white dwarfs\maybecite{2017MNRAS.468.1946H,2018AJ....156..119H}.  Similarly we mention the nearest southern magnetic white dwarf \targ{WD~0236-269}.  These objects are known to have anomalous fluxes\maybecite{2017MNRAS.468.1946H} suggesting surface features not aligned with the spin axis, as in pulsars.  It is possible that other stars with exceptional magnetic fields show such a feature.

Exoplanet systems include but are not limited to the already-mentioned \targ{HD~209458}. \targ{TOI-3884} is inferred from photometry to have both starspots and a transiting exoplanet\maybecite{2023AJ....165..249L}, and potentially there are many such targets for interferometry.  Exoplanet discovery through interferometric astrometry is also a prospect\maybecite{2025PhRvD.112h1308V}.

\section{Transients}

Transient phenomena are of course the most challenging to observe, but they are numerous.  Even confining ourselves to single-star phenomena (that is, leaving accreting and interaction systems including novae, colliding winds, and SN~Ia to another white paper) there are several kinds to explore.

Solar flares, though they may threaten civil infrastructure on Earth, are minor on the scale of the Sun.  Not so on, say, \targ{Proxima Centauri} where much stronger convection and magnetic fields produce much more prominent flares.  Spectrally intriguing features have been reported on some other red dwarfs \maybecite{2018PASJ...70...62H,2021PASJ...73...44M}, further motivating imaging of stellar flares.

Still more prominent than flares are outflows. The nearest example is \targ{$\gamma^2$~Velorum}, a Wolf-Rayet star which also happens to be in a binary.  The size of the Roche-lobe resulting from the outflow was an early interferometric measurement\maybecite{1970MNRAS.148..103H}, but mapping inhomogeneities in outflows and understanding the time variation are for future research.

Finally, there are those rarest and baddest of all single-star transients: core-collapse supernovae.   The nearest in living memory was SN~1987A, which at peak brightness had the color of \targ{Deneb} but was $\sim100$~times the size and distance. There may not be another such this century --- but if there is, you don't want to miss it.

% Microlensing \maybecite{2025ApJ...980...47M}

\section{Optical aperture synthesis beyond 10~km}

The existing and prospective observatories indicated in Figure~1 are all in the optical/infrared range but use three very different techniques, as follows.
\begin{itemize}

\item JWST is a general-purpose telescope, and in principle, its angular resolution could be improved by reshaping its mirror segments slightly and spreading them out in coordinated fashion over tens of km.  But this is far from present-day technology.

\item CHARA is a Michelson Stellar Interferometer (MSI), such as first used to measure the angular size of \targ{Betelgeuse}\cite{1921ApJ....53..249M}.  Extension to 10s of km are possible in principle, but have not yet been considered, because of the challenge of maintaining optical coherence.

\item The others are Hanbury Brown and Twiss (HBT) interferometers (also called stellar intensity interferometers) such as first used to measure the angular size of \targ{Sirius}\cite{1958RSPSA.248..222B}.  HBT uses second-order coherence, bypassing the need for classical coherence.  Extension to 10s of km has not been demonstrated yet, but is considered feasible. The SNR is set by how close one can get to the uncertainty-principle limit of simultaneously measuring both energy and arrival-time of individual photons.  Technology is still orders of magnitude away from the limit --- which has been the obstacle to wider adoption --- but is under intensive development.

\end{itemize}

These considerations suggests a network connecting all current and future telescopes across the full Paranal-Armazones area, enabling intensity interferometry across all pairs of telescopes.  This would be analogous to the existing interferometric network across the VLT and Auxiliary Telescopes, but over 100-fold longer distances using HBT interferometry.

\begin{multicols}{2}

\tolerance=1000

\def\aj{AJ\ }
\def\aap{A\&A\ }
\def\apj{ApJ\ }
\def\apjl{ApJL\ }
\def\mnras{MNRAS\ }
\def\prd{PRD\ }
\def\pasj{PASJ\ }
\def\eprint#1#2{#1:#2}

\let\oldthebibliography\thebibliography
\let\endoldthebibliography\endthebibliography
\renewenvironment{thebibliography}[1]{
  \begin{oldthebibliography}{#1}
    \setlength{\itemsep}{0em}
    \setlength{\parskip}{0em}
}
{
  \end{oldthebibliography}
}

%\begingroup
%\setlength{\bibsep}{0pt}  % Reduces spacing between references 
%--> By adding  this I needed natbib, which also changed the format of the references. Feel free to comment this and natbib out to revert. 
%\endgroup
%Thanks, found a variant without natbib, which was introducing extra ()s
\bibliographystyle{terse}
\small{\bibliography{main.bib}}

@ARTICLE{2018PASJ...70...62H,
       author = {{Honda}, Satoshi and {Notsu}, Yuta and {Namekata}, Kosuke and {Notsu}, Shota and {Maehara}, Hiroyuki and {Ikuta}, Kai and {Nogami}, Daisaku and {Shibata}, Kazunari},
        title = "{Time-resolved spectroscopic observations of an M-dwarf flare star EV Lacertae during a flare}",
      journal = {\pasj},
         year = 2018,
        month = aug,
       volume = {70},
       number = {4},
          eid = {62},
        pages = {62},
          doi = {10.1093/pasj/psy055},
archivePrefix = {arXiv},
       eprint = {1804.03771},
 primaryClass = {astro-ph.SR},
       adsurl = {https://ui.adsabs.harvard.edu/abs/2018PASJ...70...62H}
}

@ARTICLE{2021PASJ...73...44M,
       author = {{Maehara}, Hiroyuki and {Notsu}, Yuta and {Namekata}, Kousuke and {Honda}, Satoshi and {Kowalski}, Adam F. and {Katoh}, Noriyuki and {Ohshima}, Tomohito and {Iida}, Kota and {Oeda}, Motoki and {Murata}, Katsuhiro L. and {Yamanaka}, Masayuki and {Takagi}, Kengo and {Sasada}, Mahito and {Akitaya}, Hiroshi and {Ikuta}, Kai and {Okamoto}, Soshi and {Nogami}, Daisaku and {Shibata}, Kazunari},
        title = "{Time-resolved spectroscopy and photometry of M dwarf flare star YZ Canis Minoris with OISTER and TESS: Blue asymmetry in the H{\ensuremath{\alpha}} line during the non-white light flare}",
      journal = {\pasj},
         year = 2021,
        month = feb,
       volume = {73},
       number = {1},
        pages = {44-65},
          doi = {10.1093/pasj/psaa098},
archivePrefix = {arXiv},
       eprint = {2009.14412},
 primaryClass = {astro-ph.SR},
       adsurl = {https://ui.adsabs.harvard.edu/abs/2021PASJ...73...44M}
}

@ARTICLE{2023AJ....165..249L,
       author = {{Libby-Roberts}, Jessica E. and {Schutte}, Maria and {Hebb}, Leslie and {Kanodia}, Shubham and {Ca{\~n}as}, Caleb I. and {Stef{\'a}nsson}, Gu{\dh}mundur and {Lin}, Andrea S.~J. and {Mahadevan}, Suvrath and {Parts}, Winter and {Powers}, Luke and {Wisniewski}, John and {Bender}, Chad F. and {Cochran}, William D. and {Diddams}, Scott A. and {Everett}, Mark E. and {Gupta}, Arvind F. and {Halverson}, Samuel and {Kobulnicky}, Henry A. and {Kowalski}, Adam F. and {Larsen}, Alexander and {Monson}, Andrew and {Ninan}, Joe P. and {Parker}, Brock A. and {Ramsey}, Lawrence W. and {Robertson}, Paul and {Schwab}, Christian and {Swaby}, Tera N. and {Terrien}, Ryan C.},
        title = "{An In-depth Look at TOI-3884b: A Super-Neptune Transiting an M4Dwarf with Persistent Starspot Crossings}",
      journal = {\aj},
         year = 2023,
        month = jun,
       volume = {165},
       number = {6},
          eid = {249},
        pages = {249},
          doi = {10.3847/1538-3881/accc2f},
       adsurl = {https://ui.adsabs.harvard.edu/abs/2023AJ....165..249L}
}

@ARTICLE{2020MNRAS.498.4577B,
       author = {{Baumgartner}, Sandra and {Bernardini}, Mauro and {Canivete Cuissa}, Jos{\'e} R. and {de Laroussilhe}, Hugues and {Mitchell}, Alison M.~W. and {Neuenschwander}, Benno A. and {Saha}, Prasenjit and {Schaeffer}, Timoth{\'e}e and {Soyuer}, Deniz and {Zwick}, Lorenz},
        title = "{Towards a polarization prediction for LISA via intensity interferometry}",
      journal = {\mnras},
         year = 2020,
        month = nov,
       volume = {498},
       number = {3},
        pages = {4577-4589},
          doi = {10.1093/mnras/staa2638},
archivePrefix = {arXiv},
       eprint = {2008.11538},
 primaryClass = {astro-ph.IM},
       adsurl = {https://ui.adsabs.harvard.edu/abs/2020MNRAS.498.4577B}
}

@ARTICLE{2023MNRAS.524.1692F,
       author = {{Farmer}, R. and {Renzo}, M. and {G{\"o}tberg}, Y. and {Bellinger}, E. and {Justham}, S. and {de Mink}, S.~E.},
        title = "{Observational predictions for Thorne-{\.Z}ytkow objects}",
      journal = {\mnras},
         year = 2023,
        month = sep,
       volume = {524},
       number = {2},
        pages = {1692-1709},
          doi = {10.1093/mnras/stad1977},
archivePrefix = {arXiv},
       eprint = {2305.07337},
 primaryClass = {astro-ph.SR},
       adsurl = {https://ui.adsabs.harvard.edu/abs/2023MNRAS.524.1692F}
}

@ARTICLE{1921ApJ....53..249M,
       author = {{Michelson}, A.~A. and {Pease}, F.~G.},
        title = "{Measurement of the Diameter of {\ensuremath{\alpha}} Orionis with the Interferometer.}",
      journal = {\apj},
         year = 1921,
        month = may,
       volume = {53},
        pages = {249-259},
          doi = {10.1086/142603},
       adsurl = {https://ui.adsabs.harvard.edu/abs/1921ApJ....53..249M}
}

@ARTICLE{2006ApJ...645..664A,
       author = {{Aufdenberg}, J.~P. and {M{\'e}rand}, A. and {Coud{\'e} du Foresto}, V. and {Absil}, O. and {Di Folco}, E. and {Kervella}, P. and {Ridgway}, S.~T. and {Berger}, D.~H. and {ten Brummelaar}, T.~A. and {McAlister}, H.~A. and {Sturmann}, J. and {Sturmann}, L. and {Turner}, N.~H.},
        title = "{First Results from the CHARA Array. VII. Long-Baseline Interferometric Measurements of Vega Consistent with a Pole-On, Rapidly Rotating Star}",
      journal = {\apj},
         year = 2006,
        month = jul,
       volume = {645},
       number = {1},
        pages = {664-675},
          doi = {10.1086/504149},
archivePrefix = {arXiv},
       eprint = {astro-ph/0603327},
 primaryClass = {astro-ph},
       adsurl = {https://ui.adsabs.harvard.edu/abs/2006ApJ...645..664A}
}

@ARTICLE{2024ApJ...974..113A,
       author = {{Anugu}, Narsireddy and {Baron}, Fabien and {Monnier}, John D. and {Gies}, Douglas R. and {Roettenbacher}, Rachael M. and {Schaefer}, Gail H. and {Montarg{\`e}s}, Miguel and {Kraus}, Stefan and {Le Bouquin}, Jean-Baptiste and {Anderson}, Matthew D. and {ten Brummelaar}, Theo and {Codron}, Isabelle and {Farrington}, Christopher D. and {Gardner}, Tyler and {Gutierrez}, Mayra and {K{\"o}hler}, Rainer and {Lanthermann}, Cyprien and {Norris}, Ryan and {Scott}, Nicholas J. and {Setterholm}, Benjamin R. and {Vargas}, Norman L.},
        title = "{CHARA Near-infrared Imaging of the Yellow Hypergiant Star {\ensuremath{\rho}} Cassiopeiae: Convection Cells and Circumstellar Envelope}",
      journal = {\apj},
         year = 2024,
        month = oct,
       volume = {974},
       number = {1},
          eid = {113},
        pages = {113},
          doi = {10.3847/1538-4357/ad6b2b},
archivePrefix = {arXiv},
       eprint = {2408.02756},
 primaryClass = {astro-ph.SR},
       adsurl = {https://ui.adsabs.harvard.edu/abs/2024ApJ...974..113A}
}

@ARTICLE{2011ApJ...732...68C,
       author = {{Che}, X. and {Monnier}, J.~D. and {Zhao}, M. and {Pedretti}, E. and {Thureau}, N. and {M{\'e}rand}, A. and {ten Brummelaar}, T. and {McAlister}, H. and {Ridgway}, S.~T. and {Turner}, N. and {Sturmann}, J. and {Sturmann}, L.},
        title = "{Colder and Hotter: Interferometric Imaging of {\ensuremath{\beta}} Cassiopeiae and {\ensuremath{\alpha}} Leonis}",
      journal = {\apj},
         year = 2011,
        month = may,
       volume = {732},
       number = {2},
          eid = {68},
        pages = {68},
          doi = {10.1088/0004-637X/732/2/68},
archivePrefix = {arXiv},
       eprint = {1105.0740},
 primaryClass = {astro-ph.SR},
       adsurl = {https://ui.adsabs.harvard.edu/abs/2011ApJ...732...68C}
}

@ARTICLE{2021A&A...654A..19D,
       author = {{Domiciano de Souza}, A. and {Zorec}, J. and {Millour}, F. and {Le Bouquin}, J.-B. and {Spang}, A. and {Vakili}, F.},
        title = "{Refined fundamental parameters of Canopus from combined near-IR interferometry and spectral energy distribution}",
      journal = {\aap},
         year = 2021,
        month = oct,
       volume = {654},
          eid = {A19},
        pages = {A19},
          doi = {10.1051/0004-6361/202140478},
archivePrefix = {arXiv},
       eprint = {2109.07153},
 primaryClass = {astro-ph.SR},
       adsurl = {https://ui.adsabs.harvard.edu/abs/2021A&A...654A..19D}
}

@ARTICLE{1958RSPSA.248..222B,
       author = {{Hanbury Brown}, R. and {Twiss}, R.~Q.},
        title = "{Interferometry of the Intensity Fluctuations in Light. IV. A Test of an Intensity Interferometer on Sirius A}",
      journal = {Proc.\ Roy.\ Soc.\ A},
         year = 1958,
        month = nov,
       volume = {248},
       number = {1253},
        pages = {222-237},
          doi = {10.1098/rspa.1958.0240},
       adsurl = {https://ui.adsabs.harvard.edu/abs/1958RSPSA.248..222B}
}

@ARTICLE{1971MNRAS.151..161H,
       author = {{Herbison-Evans}, D. and {Hanbury Brown}, R. and {Davis}, J. and {Allen}, L.~R.},
        title = "{A study of alpha Virginis with an intensity interferometer.}",
      journal = {\mnras},
         year = 1971,
        month = jan,
       volume = {151},
        pages = {161},
          doi = {10.1093/mnras/151.2.161},
       adsurl = {https://ui.adsabs.harvard.edu/abs/1971MNRAS.151..161H}
}

@ARTICLE{1970MNRAS.148..103H,
       author = {{Hanbury Brown}, R. and {Davis}, J. and {Herbison-Evans}, D. and {Allen}, L.~R.},
        title = "{A study of {\ensuremath{\gamma}}$^{2}$ Velorum with a stellar intensity interferometer.}",
      journal = {\mnras},
         year = 1970,
        month = jan,
       volume = {148},
        pages = {103-117},
          doi = {10.1093/mnras/148.1.103},
       adsurl = {https://ui.adsabs.harvard.edu/abs/1970MNRAS.148..103H}
}

@ARTICLE{2025arXiv250615027A,
       author = {{Archer}, A. and {Aufdenberg}, J.~P. and {Bangale}, P. and {Bartkoske}, J.~T. and {Benbow}, W. and {Buckley}, J.~H. and {Chen}, Y. and {Chin}, N.~B.~Y. and {Christiansen}, J.~L. and {Chromey}, A.~J. and {Duerr}, A. and {Escobar Godoy}, M. and {Feldman}, S. and {Feng}, Q. and {Filbert}, S. and {Fortson}, L. and {Furniss}, A. and {Hanlon}, W. and {Hervet}, O. and {Hinrichs}, C.~E. and {Holder}, J. and {Hughes}, Z. and {Humensky}, T.~B. and {Jin}, W. and {Johnson}, M.~N. and {Kertzman}, M. and {Kherlakian}, M. and {Kieda}, D. and {Korzoun}, N. and {LeBohec}, T. and {Lisa}, M.~A. and {Lundy}, M. and {Maier}, G. and {Matthews}, N. and {Moriarty}, P. and {Mukherjee}, R. and {Ning}, W. and {Ong}, R.~A. and {Pandey}, A. and {Pohl}, M. and {Pueschel}, E. and {Quinn}, J. and {Rabinowitz}, P.~L. and {Ragan}, K. and {Reynolds}, P.~T. and {Ribeiro}, D. and {Roache}, E. and {Rose}, J.~G. and {Sadeh}, I. and {Saha}, L. and {Santander}, M. and {Scott}, J. and {Sembroski}, G.~H. and {Shang}, R. and {Tak}, D. and {Tucci}, J.~V. and {Valverde}, J. and {Vassiliev}, V.~V. and {Williams}, D.~A. and {Wong}, S.~L. and {The VERITAS Collaboration}},
        title = "{Measurement of the photosphere oblateness of $γ$ Cassiopeiae via Stellar Intensity Interferometry with the VERITAS Observatory}",
      journal = {},
         year = 2025,
        month = jun,
          eid = {arXiv:2506.15027},
        pages = {arXiv:2506.15027},
          doi = {10.48550/arXiv.2506.15027},
archivePrefix = {arXiv},
       eprint = {2506.15027},
 primaryClass = {astro-ph.SR},
       adsurl = {https://ui.adsabs.harvard.edu/abs/2025arXiv250615027A}
}

@ARTICLE{2022MNRAS.515....1D,
       author = {{de Almeida}, E.~S.~G. and {Hugbart}, M. and {Domiciano de Souza}, A. and {Rivet}, J.-P. and {Vakili}, F. and {Siciak}, A. and {Labeyrie}, G. and {Garde}, O. and {Matthews}, N. and {Lai}, O. and {Vernet}, D. and {Kaiser}, R. and {Guerin}, W.},
        title = "{Combined spectroscopy and intensity interferometry to determine the distances of the blue supergiants P Cygni and Rigel}",
      journal = {\mnras},
         year = 2022,
        month = sep,
       volume = {515},
       number = {1},
        pages = {1-12},
          doi = {10.1093/mnras/stac1617},
archivePrefix = {arXiv},
       eprint = {2204.00372},
 primaryClass = {astro-ph.SR},
       adsurl = {https://ui.adsabs.harvard.edu/abs/2022MNRAS.515....1D}
}

@ARTICLE{2025PhRvD.112h1308V,
       author = {{Van Tilburg}, Ken and {Baryakhtar}, Masha and {Galanis}, Marios and {Weiner}, Neal},
        title = "{Astrometry with extended-path intensity correlation}",
      journal = {\prd},
         year = 2025,
        month = oct,
       volume = {112},
       number = {8},
          eid = {L081308},
        pages = {L081308},
          doi = {10.1103/gx2r-38yz},
archivePrefix = {arXiv},
       eprint = {2307.03221},
 primaryClass = {astro-ph.IM},
       adsurl = {https://ui.adsabs.harvard.edu/abs/2025PhRvD.112h1308V}
}

@ARTICLE{2017NatAs...1..690C,
       author = {{Cotton}, Daniel V. and {Bailey}, Jeremy and {Howarth}, Ian D. and {Bott}, Kimberly and {Kedziora-Chudczer}, Lucyna and {Lucas}, P.~W. and {Hough}, J.~H.},
        title = "{Polarization due to rotational distortion in the bright star Regulus}",
      journal = {Nature Astronomy},
         year = 2017,
        month = sep,
       volume = {1},
        pages = {690-696},
          doi = {10.1038/s41550-017-0238-6},
archivePrefix = {arXiv},
       eprint = {1804.06576},
 primaryClass = {astro-ph.SR},
       adsurl = {https://ui.adsabs.harvard.edu/abs/2017NatAs...1..690C}
}

@ARTICLE{1975ApJ...195..137S,
       author = {{Schwarzschild}, M.},
        title = "{On the scale of photospheric convection in red giants and supergiants.}",
      journal = {\apj},
         year = 1975,
        month = jan,
       volume = {195},
        pages = {137-144},
          doi = {10.1086/153313},
       adsurl = {https://ui.adsabs.harvard.edu/abs/1975ApJ...195..137S}
}

@ARTICLE{2018MNRAS.481.2361J,
       author = {{Joyce}, S.~R.~G. and {Barstow}, M.~A. and {Holberg}, J.~B. and {Bond}, H.~E. and {Casewell}, S.~L. and {Burleigh}, M.~R.},
        title = "{The gravitational redshift of Sirius B}",
      journal = {\mnras},
         year = 2018,
        month = dec,
       volume = {481},
       number = {2},
        pages = {2361-2370},
          doi = {10.1093/mnras/sty2404},
archivePrefix = {arXiv},
       eprint = {1809.01240},
 primaryClass = {astro-ph.SR},
       adsurl = {https://ui.adsabs.harvard.edu/abs/2018MNRAS.481.2361J}
}

@ARTICLE{2017ApJ...848...11B,
       author = {{B{\'e}dard}, A. and {Bergeron}, P. and {Fontaine}, G.},
        title = "{Measurements of Physical Parameters of White Dwarfs: A Test of the Mass-Radius Relation}",
      journal = {\apj},
         year = 2017,
        month = oct,
       volume = {848},
       number = {1},
          eid = {11},
        pages = {11},
          doi = {10.3847/1538-4357/aa8bb6},
archivePrefix = {arXiv},
       eprint = {1709.02324},
 primaryClass = {astro-ph.SR},
       adsurl = {https://ui.adsabs.harvard.edu/abs/2017ApJ...848...11B}
}

@ARTICLE{2017Sci...356.1046S,
       author = {{Sahu}, Kailash C. and {Anderson}, Jay and {Casertano}, Stefano and {Bond}, Howard E. and {Bergeron}, Pierre and {Nelan}, Edmund P. and {Pueyo}, Laurent and {Brown}, Thomas M. and {Bellini}, Andrea and {Levay}, Zoltan G. and {Sokol}, Joshua and {Dominik}, Martin and {Calamida}, Annalisa and {Kains}, No{\'e} and {Livio}, Mario},
        title = "{Relativistic deflection of background starlight measures the mass of a nearby white dwarf star}",
      journal = {Science},
         year = 2017,
        month = jun,
       volume = {356},
       number = {6342},
        pages = {1046-1050},
          doi = {10.1126/science.aal2879},
archivePrefix = {arXiv},
       eprint = {1706.02037},
 primaryClass = {astro-ph.SR},
       adsurl = {https://ui.adsabs.harvard.edu/abs/2017Sci...356.1046S}
}

\end{multicols}

\mayberef

\end{document}